\documentclass[10pt,aps,prl,reprint,floatfix,floats,showpacs,superscriptaddress,raggedbottom]{revtex4-1}
\usepackage{graphicx,latexsym}
\usepackage{dcolumn}
\usepackage{amssymb,amsmath,bm}
\pdfoutput=1

\begin{document}


\title{Signature of snaking states in the conductance of core-shell nanowires}

\author{Tomas Orn Rosdahl}
\email{torosdahl@gmail.com}
\affiliation{School of Science and Engineering, Reykjavik University, Menntavegur 1, IS-101 Reykjavik, Iceland}
\author{Andrei Manolescu}
\affiliation{School of Science and Engineering, Reykjavik University, Menntavegur 1, IS-101 Reykjavik, Iceland}
\author{Vidar Gudmundsson}
\affiliation{Science Institute, University of Iceland, Dunhaga 3,
         IS-107 Reykjavik, Iceland}

\begin{abstract}
We model a core-shell nanowire (CSN) by a cylindrical surface of finite length.
A uniform magnetic field perpendicular to the axis of the cylinder forms
electron states along the lines of zero radial field projection, which
can classically be described as snaking states. In a strong field, these
states converge pairwise to quasidegenerate levels, which are situated at
the bottom of the energy spectrum. We calculate the conductance of the
CSN by coupling it to leads, and predict that the snaking states govern
transport at low chemical potential, forming isolated peaks, each of
which may be split in two by applying a transverse electric field. If
the contacts with the leads do not completely surround the CSN, as is
usually the case in experiments, the amplitude of the snaking peaks
changes when the magnetic field is rotated, determined by the overlap
of the contacts with the snaking states.  \end{abstract}


\maketitle

\emph{Introduction.}\----The transport properties of semiconductor
nanostructures subjected to an external magnetic field remains a topic
of intense research to this day. A
recently conceived heterostructure is the core-shell nanowire (CSN)
\cite{Li2006,Thelander2006,Jung2008,Wong2011}. It consists of a thin layer
(shell) surrounding a core in a tubular geometry, which may be engineered
such that carriers are confined to the shell, e.\ g.\ a conductive InAs
shell surrounding a GaAs core \cite{Rieger2012}. By piercing them with a
longitudinal magnetic flux, CSNs have been applied to study the Aharonov-Bohm effect
\cite{Tserkovnyak2006,Gladilin2013,Rosdahl2014}, which manifests as flux-periodic oscillations
in e.\ g.\ magnetoconductance \cite{Jung2008,Gul2014}.

Fascinating physics also manifests if a perpendicular magnetic
field is applied. In this case, orbital electron motion is governed by the radial
field projection, such that the magnetic field becomes effectively
inhomogeneous. Hence, such systems are comparable to the older topic of
a planar electron gas in an inhomogeneous magnetic field
\cite{Muller1992,Ibrahim1995,Ye1995,Zwerschke1999}.
In Ref.\ \onlinecite{Friedland2007}, transport measurements were performed
on segments of cylinders, and a notable asymmetry in the magnetoresistance
was observed with respect to field orientation. Furthermore, simulations of a
cylindrical electron gas immersed in a perpendicular field predicted the
formation of states with distinct characteristics, localized in different regions
around the circumference \cite{Ferrari2008,Snake}. Propagating states
tend to be localized in narrow channels, where the magnetic field is
parallel to the surface, laterally confined by a Lorentz force. On
the other hand, states with small wave numbers are condensed into a
highly-degenerate Landau level, localized in regions where the field
and surface are perpendicular, similar to a planar two-dimensional
electron gas (2DEG). In Ref.\ \onlinecite{Bellucci2010}, a semiclassical
analysis was applied to identify these states as snaking and cyclotron
states, respectively. More recent work saw the simulation of CSNs with
hexagonal cross sections, which is more realistic than a circular one. The
same qualitative behavior was observed, but with added localization
effects due to the corners \cite{Ferrari2009,Royo2013}. Indeed, in Ref.\
\onlinecite{Royo2013}, magnetoconductance calculations are presented,
but to our knowledge, the manifestations of snaking and cyclotron states
in the transport characteristics of CSNs have not been reported.

In this paper, we calculate the electron states on a cylindrical surface
of finite length, immersed in a perpendicular magnetic field, and analyze
the transport signatures of said states by coupling the cylinder to leads
using a Green's function method. Emphasis is on the snaking states and how
they manifest in transport. We probe the experimentally relevant effects of a
nonzero shell thickness, and of nonuniform coupling by allowing the contacts
with the leads to connect to the cylinder over restricted angular intervals,
and/or to its interior. At large magnetic fields, the states
essentially fall into three categories, each located in a different part
of the cylinder, namely cyclotron states condensed into a Landau level,
edge states localized at the cylinder ends, and snaking states, which
are the energetically lowest ones. The transport signature of each state type
is interpreted in terms of its distribution in the finite system and
overlap with the contacts. In particular, there is a visible asymmetry
in the snaking conductance peaks, which appear at low chemical potential, depending
on whether the contacts overlap with the snaking states or not, which
in practice can be changed by rotating the magnetic field.
\begin{figure}[htbq]
\includegraphics[width=0.35\textwidth,angle=0]{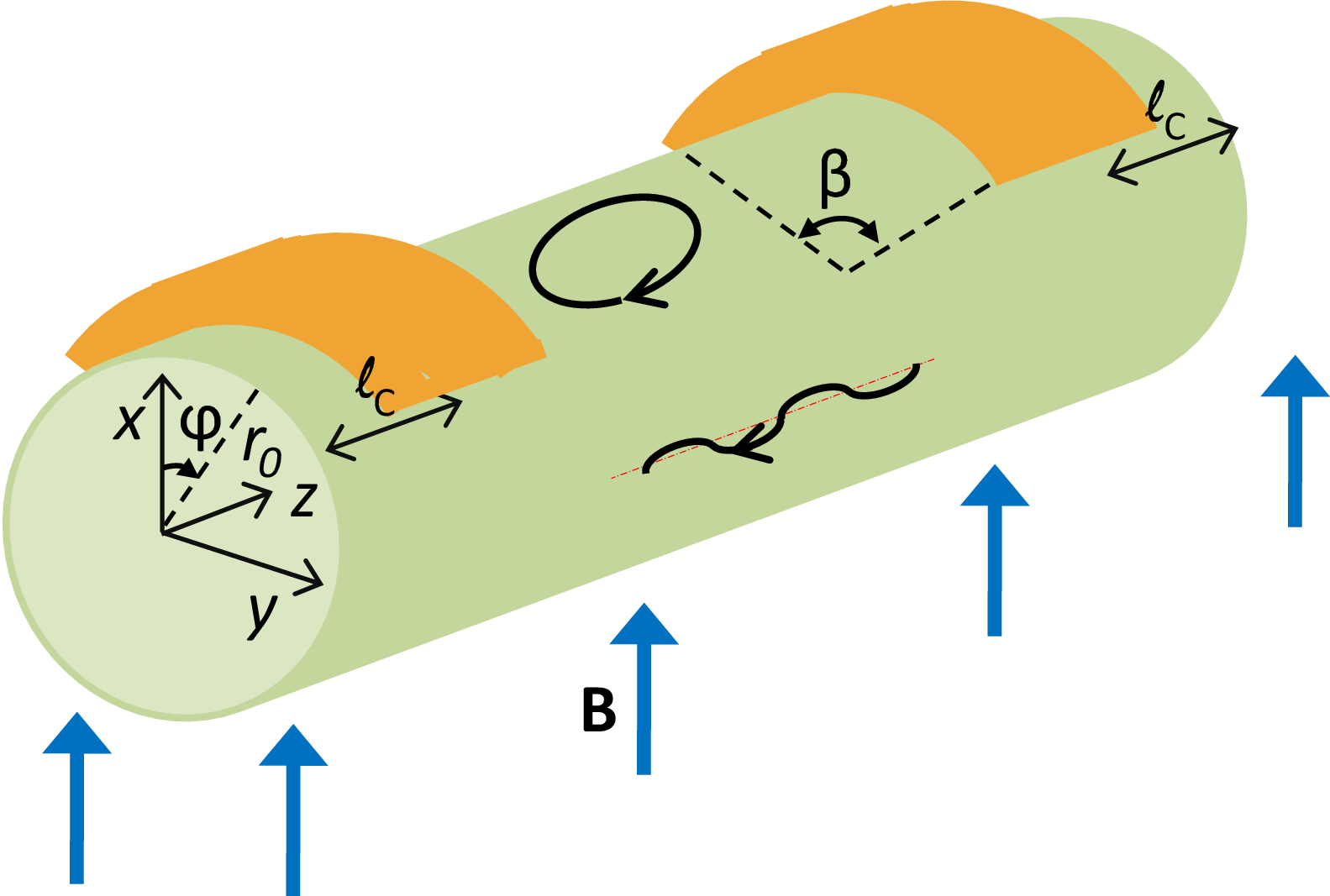} 
\caption{A model cylindrical CSN in a perpendicular
magnetic field $\boldsymbol{B}$ (blue arrows). Snaking states (wavy
arrow) form around the lines of zero radial field projection (dotted
thin red line). Cyclotron states (circular arrow) form where the
radial projection is maximum. Two contacts (gold) with conducting leads
(not shown) are placed on the nanowire over an angular interval $\beta$,
centered at distances $\ell_C$ from the ends of the nanowire.}
\label{fig:sample}
\end{figure}

\emph{Model.}\----The system under consideration consists of
noninteracting electrons confined to a cylindrical surface of radius $r_0$
and length $L_0$, narrow enough to be treated as infinitely thin, with
hard-wall boundary conditions imposed at the edges. 
In a transverse magnetic field $\boldsymbol{B} = B\hat{\mathbf{x}}$,
electron orbital motion is governed by the radial projection
$\boldsymbol{B}\cdot\hat{\mathbf{r}}$. It is nonuniform around the cylinder
circumference, breaking the rotational symmetry of the system, as is clear
from the vector potential $\boldsymbol{A} = r_0B\sin{(\varphi)}\hat{\mathbf{z}}$.
Figure \ref{fig:sample} shows a sketch of the system.
The single-electron Hamiltonian is
\begin{equation}
H_C = H_O + H_Z, \label{eq:HC} 
\end{equation}
where the orbital part $H_O$ is the kinetic energy term
\begin{equation}
H_O = \frac{p_{\varphi}^2 + p_z^2}{2m_e} 
+ r_0 \omega_c \sin{(\varphi)}p_z 
+\frac{\hbar \omega_c}{2} \frac{r_0^2}{(l_B)^2}\sin^2{(\varphi)}, 
\label{eq:HO} 
\end{equation}
with $l_B^2 = \hbar/eB$, $\omega_c = eB/m_e$ and the effective electron mass $m_e$. 
Electron spin is included through the Zeeman term
\begin{equation} 
H_Z = \frac{\hbar \omega_c g_e m_e}{4m_0}\sigma_x, 
\label{eq:HZ} 
\end{equation}
where $g_e$ is the effective electron g-factor and $m_0$ the free
electron mass. The orbital part of the time-independent
Schr\"{o}dinger equation, $H_C | a \rangle = \epsilon^C_a | a \rangle$,
is diagonalized in the basis of eigenstates of $H_O$ with $B = 0$ T.

To evaluate the transport properties of the finite cylinder, we couple
it to two leads via contacts, shown schematically in Fig.\ \ref{fig:sample},
and apply a Green's function formalism to calculate the
phase-coherent conductance $G$ in the linear-response regime
\cite{Datta,Brandbygge2002,Thygesen2003,Tada2004,Paulsson2008,Rosdahl2014}.
It follows from the spin-resolved Landauer formula
\begin{equation} 
G(\mu) = -\frac{e^2}{h} \int \frac{\partial f}{\partial E} \mathrm{Tr}
\left[G_C^{\dagger}\Gamma_R G_C \Gamma_L \right] \mathrm{d}E, 
\label{eq:G} 
\end{equation}
where $f(E,\mu,T)$ is the Fermi-Dirac function, evaluated at the chemical
potential $\mu$ and temperature $T$, both of which are assumed constant
throughout the coupled system. Here, $G_C^{-1}(E) = E-H_C-\Sigma_L
- \Sigma_R$ is the inverse retarded Green's operator of the coupled central
cylinder, where the left ($L$) and right ($R$) leads enter through the
self-energies $\Sigma_j$, which are generally complex and result in
level broadening in the central part, characterized by the operators
$\Gamma_j = i(\Sigma_j - \Sigma_j^{\dagger} )$ \cite{Datta,Kurth2005}. The operators in Eq.\
(\ref{eq:G}) act on the central system subspace and are constructed in
the basis of eigenstates of $H_C$ [Eq.\ (\ref{eq:HC})]. By assuming
coupling terms between the central part and the leads at each junction,
the self-energies assume the form
\begin{equation} 
\Sigma_j = H_{jC}^{\dagger}g_j H_{jC}, \hspace{0.2 cm} j = L,R, 
\label{eq:Sigma} 
\end{equation}
where $H_{jC}$ couples lead $j$ to the finite cylinder and $g_j^{-1}(E)
= E-H_j+i\eta$ is the inverse retarded Green's operator of the uncoupled lead,
described by the Hamiltonian $H_j$. While states in the leads and the
central part have no spatial overlap, the transmission of electrons
is facilitated through the coupling terms in the self-energy matrix
elements by introducing a nonlocal coupling kernel, which in turn models
the contacts. Applying in Eq.\ (\ref{eq:Sigma}) the closure relation for
the isolated lead eigenstates, which satisfy $H_j | q_j \rangle = \epsilon^j_{q_j} | q_j \rangle$,
one obtains the matrix element
\begin{equation} 
\langle a | \Sigma_j | b \rangle = \sum\limits_{q_j} 
\frac{\langle a | H_{jC}^{\dagger} | q_j \rangle \langle q_j | H_{jC} | b \rangle}
{E-\epsilon^{j}_{q_j}+i\eta}. 
\end{equation}
In coordinate representation, the overlap matrix elements are
[$\langle \boldsymbol{r} | a \rangle = \Psi^C_{a}(\boldsymbol{r})$, $\langle \boldsymbol{r}' | q_j \rangle = \Psi^j_{q_j}(\boldsymbol{r}')$]
\begin{equation} 
\langle a | H_{jC}^{\dagger} | b \rangle = \int_{C} \mathrm{d}^3r \int_{j} \mathrm{d}^3r' \left[\Psi^C_{a}(\boldsymbol{r}) \right]^{\dagger} K_j(\boldsymbol{r},\boldsymbol{r}') \Psi^j_{q_j}(\boldsymbol{r}'), 
\end{equation} 
where the two spatial integrals, which extend over the finite
cylinder ($C$) and lead $j$, couple through the kernel $\langle
\boldsymbol{r} | H_{jC}^{\dagger} | \boldsymbol{r}' \rangle =
K_j(\boldsymbol{r},\boldsymbol{r}')$. We use a kernel which decays
exponentially with distance from the junction, and may couple the leads
to the central part uniformly over the circumference or over restricted
angular intervals $\beta$ (Fig.\ \ref{fig:sample}), the latter of
which models the nonuniform application of contacts in experiment. A
possible shift $\ell_C$ of the kernel away from the cylinder edges is
also included, to model the application of contacts to the interior, as
is typical in experiment \cite{Jung2008,Gul2014}. In this case, only
the central system part of the kernel is shifted, leaving the kernel
overlap with states in the leads unchanged. Technically, the leads are
taken as semi-infinite cylindrical continuations of the central part 
with hard-wall boundary conditions imposed at the junctions with the
finite system \cite{Rosdahl2014}.
However, we limit ourselves to weak coupling, which ensures that the particular form
of the kernel and the leads is not paramount, as $G$ should be governed by the properties
of the finite cylinder itself, i.\ e.\ by $H_C$ in Eq.\ (\ref{eq:HC}). The present
model for the contacts has been used to qualitatively replicate experimental results \cite{Rosdahl2014}.
Alternative models, e.\ g.\ Ref.\ \onlinecite{Gudmundsson2009}, would only result
in a minor change in level broadening.

\emph{Results and discussion.}\---- Let $r_0 = 17$ nm and $L_0 = 170$
nm for an aspect ratio $L_0/r_0 = 10$, such that multiple axial
levels enter the spectrum per angular mode.
Material parameters are taken for InAs, i.\ e.\ $m_e = 0.023m_0$ and
$g_e = -14.9$. In this work, we neglect the effects of surface defects and impurities.
This is a reasonable approximation, because the measured mean free path
of InAs surface inversion layers is of the order $150$ nm, and the cylinder
under consideration is therefore close to the ballistic limit \cite{Matsuyama2000,Chuang2013}. From a
semiclassical point of view \cite{Muller1992,Bellucci2010}, lateral
confinement by the Lorentz force is expected to produce snaking electron
states where the radial field projection vanishes ($\varphi = \pm \pi/2$),
while cyclotron states condensed into a Landau level are expected
where the radial projection is maximum ($\varphi = 0$ and $\varphi
= \pi$). In these regions, $\boldsymbol{B}$ is parallel and perpendicular
to the cylinder surface, respectively, and we refer to them as the snaking
and cyclotron regions. The two types of states become localized in their
respective regions and clearly discernible if $l_B << r_0$.

\begin{figure}[htbq]
\includegraphics[width=0.48\textwidth,angle=0]{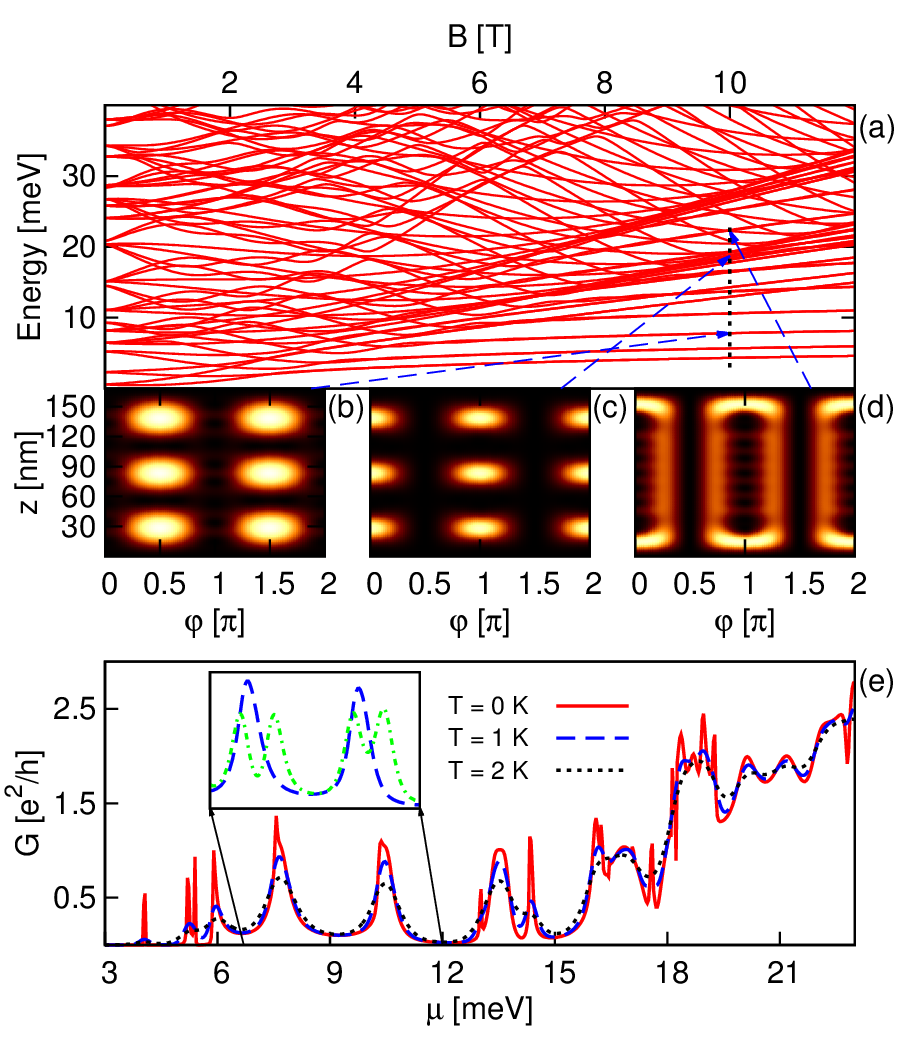}
\caption{(Color online) (a) Spectrum of the cylinder, uncoupled to leads,
as a function of $B$. At
large $B$, three distinct types of states are discernible:
snaking and cyclotron states, localized in regions where $\boldsymbol{B}$
is parallel or perpendicular to the cylinder surface, respectively,
and edge states, localized at the cylinder ends. Examples of each type
are shown in (b), (c) and (d), respectively, where bright colors
indicate high probability densities on the cylindrical surface. 
(e) Conductance of the
coupled system as a function of $\mu$, evaluated at various temperatures
with $B = 10$ T. The corresponding energy range is marked with a vertical
dotted line in (a). The low-energy peaks correspond to snaking states,
which are quasidegenerate levels in the closed system. The inset shows an example
of how a weak transverse electric field lifts this degeneracy,
splitting the snaking conductance peaks in two (dash-dotted line).}
\label{fig:1}
\end{figure}
Figure \ref{fig:1}(a) shows the spectrum of the finite cylinder, uncoupled
to leads, as a function of $B$. At large $B$ (e.\ g.\
$10$ T), the electron states roughly fall into three categories. The
energetically lowest states converge pairwise into quasidegenerate levels, all
of which are localized in the snaking regions, as exemplified in Fig.\
\ref{fig:1}(b), which shows the probability density of one such state.
Hence, we identify them as snaking states.
The quasidegeneracy arises due to the symmetry between the snaking
regions, as the magnetic confinement does not distinguish between them.
The energy difference between the quasidegenerate snaking states is very small,
unresolvable along the vertical dotted line in Fig.\ \ref{fig:1}(a),
and it tends to vanish with increasing magnetic field strength.
The quasidegeneracy can be lifted e.\ g.\ by applying a transverse electric
field, to explicitly break the symmetry between the snaking regions.
A similar, but exact, degeneracy arises on a CSN of infinite length,
where the degenerate states correspond to longitudinal propagation
in opposite directions along the cylinder axis \cite{Snake}.
Moving up in energy in the spectrum, the density of states rises
abruptly in a narrow energy interval. The relevant states are localized
in the cyclotron regions, as the example in Fig.\ \ref{fig:1}(c)
illustrates. Accordingly, these are cyclotron states condensed into
a Landau level \cite{Ferrari2008,Ferrari2009,Snake}. Observe that
the Landau level is split into spin up and down branches by the Zeeman
term. If spin splitting is neglected, its energy approximately coincides
with the lowest Landau level of a planar 2DEG, $\hbar \omega_c/2$. The
final category consists of states that are all primarily localized at the
cylinder ends, as the example in Fig.\ \ref{fig:1}(d) shows, and hence
we refer to them as edge states. Their energies are woven into a net-like
structure, reminiscent of the flux-periodic oscillations that arise when
wave functions enclose a magnetic flux \cite{Aharonov1959,Byers1961}.
Here, the net-like structure manifests because the electron wave
functions enclose a magnetic flux perpendicular to the curved cylinder
surface. Lastly, we mention that if Zeeman splitting is neglected, the
aforementioned categories become segregated in energy, such that the
snaking states are energetically lowest, followed by the Landau level,
and then by the edge states. Here, this separation is lost because of the
large g-factor in InAs, but it would still be present in a material with
a smaller g-factor, e.\ g.\ GaAs. Regardless, the energetically-lowest
states are still snaking states, with the spin down branch setting in
around $14$ meV at $B = 10$ T, among the snaking states of the spin
up branch.

The various state categories previously discussed manifest differently
in transport. In Fig.\ \ref{fig:1}(e), we show $G$ [Eq.\ (\ref{eq:G})]
with $B=10$ T as a function of $\mu$ at various temperatures. The
corresponding energy range is marked with a vertical dotted line in
Fig.\ \ref{fig:1}(a). Roughly, $G$ peaks when $\mu$ intersects with an
energy level of the finite cylinder, with level broadening and energy
shift arising due to the self-energies $\Sigma_j$, both of which are small
($\approx 1$ meV) because the systems are weakly coupled. Another
effect is thermal broadening due to smearing of the Fermi function, as
the conductance curves illustrate. Sweeping through the values of $\mu$,
we see a number of conductance peaks at low energies corresponding to
the snaking states in the finite cylinder ($3 \lesssim \mu \lesssim 17$
meV). They vary in size and shape because the corresponding states on
the finite cylinder couple differently to the contact kernel, resulting
in different self-energy matrix elements, and hence level broadening. Here,
$\ell_C = 0$ nm, and the contacts are applied uniformly over the cylinder
circumference ($\beta=2\pi$ in Fig.\ \ref{fig:sample}), so it is primarily
the difference in the states' longitudinal distribution that causes
different broadening. For example, the first peak ($\mu \approx 4$ meV)
corresponds to a state which is localized around the cylinder center.
Hence, it has significantly less overlap with the contact
kernel, which is localized at the edges, than the state corresponding
to the peak around $\mu \approx 7-8$ meV, the distribution of which
is similar to the one in Fig.\ \ref{fig:1}(b). 
As discussed in the preceding paragraph, each snaking conductance
peak arises due to a pair of quasidegenerate snaking states in the closed system.
In Fig.\ \ref{fig:1}(e), the inset shows how two of the snaking peaks ($T = 1$ K)
are affected by a weak electric field of strength $30$ $\mu$eV/nm,
oriented along the $y$ axis. This breaks the symmetry between the snaking regions,
splitting the quasidegenerate snaking states in energy, which in turn splits each conductance
peak in two by an amount proportional to the electric field strength, as the
dash-dotted curve shows. Around $\mu \approx 18$ meV, the spin up Landau level
sets in, resulting in a sharp increase in $G$. The Landau level
is furthermore characterized by a sequence of narrow conductance peaks at $T = 0$ K,
each corresponding to a cyclotron state, which are washed out and combined
by thermal broadening at $1$ and $2$ K. Above the Landau level ($\mu \gtrsim 19$
meV), we observe a number of peaks that are significantly more broadened
than the lower-energy peaks, even at $0$ K. These correspond to the
edge states in the closed system, which are primarily localized at the
cylinder ends [e.\ g.\ Fig.\ \ref{fig:1}(c)]. Accordingly, such states
overlap more with the contacts than the other types, and therefore
their broadening is more substantial.

The assumption of an infinitely narrow surface is an idealization of the shell,
which in reality has some thickness. To gauge the validity of our model, let us consider
a shell with nonvanishing thickness $d = r_2 - r_1$. The kinetic energy term is the same as in
Eq.\ (\ref{eq:HO}), but with $r_0$ replaced by the radial coordinate $r$, and the
added term $p_r^2/2m_e - \hbar^2/8m_er^2$. Here, $p_r = (\hat{\mathbf{r}}\cdot
\boldsymbol{p} + \boldsymbol{p}\cdot \hat{\mathbf{r}})/2 = -i\hbar(\partial_r + 1/2r)$
is the canonical radial momentum operator, with $\boldsymbol{p} = (p_x,p_y,p_z)$.
Imposing hard-wall boundary conditions at $r_1$ and $r_2$, we diagonalize the radial part
of the Hamiltonian in the basis of eigenstates of $p_r^2/2m_e$, namely
\begin{equation} \sqrt{\frac{2}{d}} \frac{1}{\sqrt{r}}\sin{\left( \frac{n\pi}{d}[r-r_1 ] \right)}, \hspace{0.2 cm} n \in \mathbb{Z}_{+}, \label{eq:RadBas} \end{equation}
whose eigenvalues are the familiar infinite square well energies.
The thickness is implemented in the leads analogously. Intuitively, an infinitely
narrow surface should be a good approximation if the lowest radial mode in Eq.\ (\ref{eq:RadBas})
dominates in the energy range of interest, namely $\lesssim \hbar \omega_c$.
We performed calculations with $d$ chosen such that the mean
radius $r_m = (r_1+r_2)/2$ is equivalent to $r_0$ given above. Our results (not shown here) with $d = 7$ nm
and $d = 17$ nm show the same qualitative behaviour as observed in Fig.\ \ref{fig:1}.
For the former, $d/r_m \approx 0.4$, which is realistic from an experimental point of view \cite{Gul2014,Wenz2014}.
In both cases, the energy difference between the two lowest states in Eq.\ (\ref{eq:RadBas}) is much
larger than $\hbar \omega_c$ ($50$ meV at $10$ T), such that the lowest radial mode is dominant. Indeed,
the relevant states are all localized around $r = r_m$, and snaking and cyclotron states remain spatially separated.
Furthermore, the energy spectra are nearly indistinguishable from the case with an infinitely narrow shell, and
although there are minor differences in the finer structure of individual conductance peaks,
the same trends as in Fig.\ \ref{fig:1}(e) are present, namely isolated snaking peaks followed
by the onset of the spin split Landau level.

\begin{figure}[htbq]
\includegraphics[width=0.48\textwidth,angle=0]{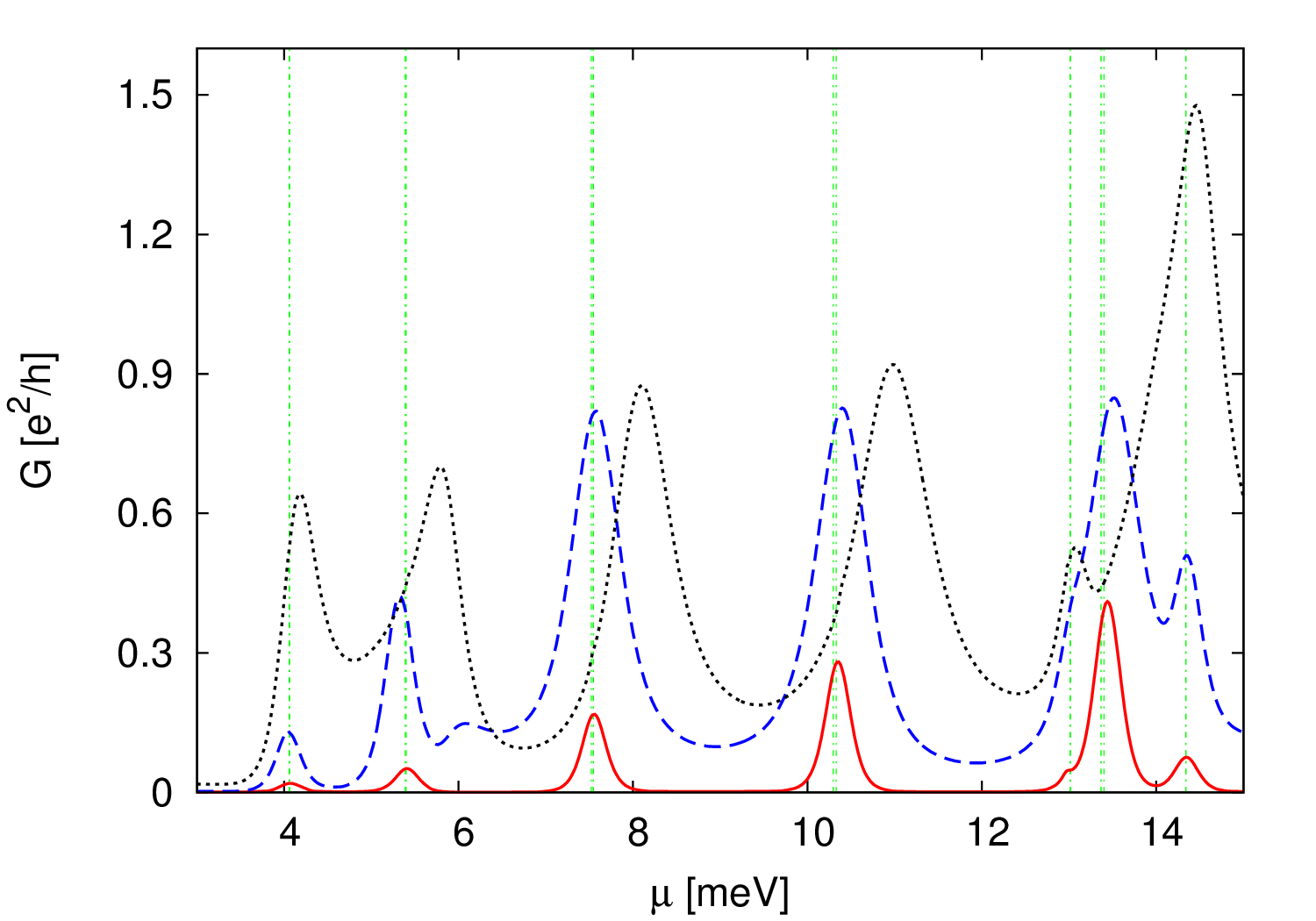}
\caption{(Color online) Conductance of the coupled system as a function
of $\mu$ with $B = 10$ T, over an energy range corresponding to the snaking states in
the spectrum [see Fig.\ \ref{fig:1}(a)]. The contacts cover
an angular interval $\beta=\pi/2$, positioned over a cyclotron region (solid line)
or a snaking region (dashed line), with $\ell_C = 0$ nm in both cases.
The conductance peaks are more pronounced in the latter case than in the former,
because the snaking states, whose energies are marked with vertical dash-dotted lines, 
have greater overlap with the contacts.
The dotted curve is obtained with the same parameters as the dashed curve,
but with the contacts shifted by $\ell_C = 11$ nm.}
\label{fig:3}
\end{figure}
The transport signature of a particular state is primarily determined by
its overlap with the coupling kernel. Furthermore, snaking, cyclotron, and
edge states are all principally localized in their respective regions
on the finite cylinder. Combining this with a restricted application
of the contacts, over limited parts of the circumference, we can tailor the transport
signatures of the various types of states. Figure \ref{fig:3} shows the
conductance of the coupled system as a function of $\mu$ with $B = 10$
T and $T = 1$ K. The $\mu$ values are chosen such that the conductance
peaks correspond to snaking states, whose energies in the closed system
are marked with vertical dash-dotted lines [see also Fig.\ \ref{fig:1}(a)].
Both leads are coupled identically to the central part with $\ell_C = 0$ nm,
over an angular interval $\beta=\pi/2$, centered at a cyclotron region (solid line) or a
snaking region (dashed line). By parity invariance, it does not matter to
which of the two snaking or cyclotron regions the contacts are applied.
The conductance peaks shown are significantly more pronounced in the latter
case than in the former, because the snaking states' overlap with the coupling
kernel is greater when the contacts connect to a snaking region than when they
connect to a cyclotron region, resulting in a visible asymmetry between the two
cases. Similarly, the cyclotron states are more prominent if the
contacts are applied to a cyclotron region (not shown).

Typically, contacts are not applied to the sample edges as we have assumed,
but to the interior \cite{Jung2008,Gul2014}. In Fig.\ \ref{fig:3},
the dotted curve shows $G$ obtained with $\ell_C = 11$ nm,
using contacts placed over a snaking region, and the same parameters as
in the preceding paragraph. Intuitively, the increased overlap with states in the
central part should yield larger self energies, resulting in more prominent level
broadening and energy shift. Indeed, the peaks are more broadened
and shifted, as compared to the dashed curve, which is identical but with $\ell_C = 0$ nm.
Nevertheless, the two curves exhibit the same qualitative behavior.

\emph{Conclusions.}\---- We have performed calculations for electrons on a
cylindrical surface, to simulate a CSN, and seen the formation of snaking,
cyclotron, and edge states in a large, perpendicular magnetic field. Each
state type has a different transport signature, with broadening primarily
governed by its overlap with the contacts. In particular, the snaking states
manifest as peaks at low chemical potentials. Each snaking peak can be
split in two by breaking the symmetry between the two snaking regions,
e.\ g.\ with a transverse electric field. Such a field can be created
with lateral gates attached to the CSN. However, in a realistic system,
it is possible that nonuniform application of contacts around the nanowire 
could induce intrinsic fields,
which might break the symmetry between the snaking regions, splitting the
snaking conductance peaks, even in the absence of a transverse electric field.
Regardless, a weak electric field might still be used to tune the spacing between
the peaks, possibly even to recombine them. Our results hold even in the
presence of a nonzero shell thickness. We have furthermore shown that
if the contacts only connect to restricted parts of the nanowire, the snaking conductance peaks
exhibit a pronounced asymmetry with varying contact placement along the
circumference, depending on the snaking states' overlap with the
contacts. Moving the contacts in such a manner relative to the
snaking regions may be achieved experimentally by rotating
the magnetic field in the plane perpendicular to the axis of the cylinder.

\begin{acknowledgments}
We thank Thomas Sch\"apers for enlightening discussions. This work was supported
by the Research Fund of the University of Iceland and the Icelandic Research and Instruments Funds.
\end{acknowledgments}

%
%
%
\bibliographystyle{apsrev4-1}
\bibliography{citations}

\end{document}